# Iterative Learning Control-Informed Reinforcement Learning for Batch Process Control

Runze Lin [1,♣], Ziqi Zhuo [1,♣], Junghui Chen [2,*], Lei Xie [1,*], Hongye Su [1]

[1] State Key Laboratory of Industrial Control Technology, Institute of Cyber-Systems and Control, Zhejiang University, Hangzhou 310027, China

[2] Department of Chemical Engineering, Chung-Yuan Christian University, Taoyuan 32023, Taiwan, R.O.C.

♣ The two authors contributed equally to this work.

* Corresponding authors.

**Abstract:** A significant limitation of Deep Reinforcement Learning (DRL) is the stochastic uncertainty in actions generated during exploration-exploitation, which poses substantial safety risks during both training and deployment. In industrial process control, the lack of formal stability and convergence guarantees further inhibits adoption of DRL methods by practitioners. Conversely, Iterative Learning Control (ILC) represents a well-established autonomous control methodology for repetitive systems, particularly in batch process optimization. ILC achieves desired control performance through iterative refinement of control laws, either between consecutive batches or within individual batches, to compensate for both repetitive and non-repetitive disturbances. This study introduces an Iterative Learning Control-Informed Reinforcement Learning (IL-CIRL) framework for training DRL controllers in dual-layer batch-to-batch and within-batch control architectures for batch processes. The proposed method incorporates Kalman filter-based state estimation within the iterative learning structure to guide DRL agents toward control policies that satisfy operational constraints and ensure stability guarantees. This approach enables the systematic design of DRL controllers for batch processes operating under multiple disturbance conditions.

# 1 Introduction

In the process industry, batch manufacturing represents a critical production methodology widely deployed across sectors vital to economic development and public welfare, including chemicals, pharmaceuticals, food processing, and materials synthesis [1, 2]. These processes constitute essential technical infrastructure for producing high-value-added, multi-variety, small-batch products. Applications such as specialty chemical preparation, antibiotic manufacturing, functional food processing, and advanced composite material synthesis depend extensively on batch processes to meet demanding product specifications and quality requirements. Despite their importance, batch process optimization presents significant technical challenges. Unlike continuous processes, batch systems require control architectures capable of managing nonlinear dynamics and time-varying behaviors, optimizing production trajectories across successive batches, and maintaining precise trajectory tracking within individual batches [3]. Consequently, batch process optimization encompasses two distinct control problems: inter-batch process optimization and intra-batch precision control. Developing robust control strategies that effectively handle disturbances both between batches (batch-to-batch) and within individual batches (within-batch) while ensuring system stability and efficiency across varying production conditions represents a fundamental challenge in this field.

Traditional batch process control methods, such as Model Predictive Control (MPC) and Iterative Learning Control (ILC), have demonstrated significant effectiveness in industrial applications [4-13]. MPC optimizes batch process performance by solving model-based optimal control problems; however, its reliance on precise mathematical models limits its applicability to highly nonlinear or complex systems [14, 15]. Conversely, ILC enhances control performance in repetitive tasks by iteratively learning from previous batch executions to mitigate process disturbances and converge toward desired trajectories. Nevertheless, conventional ILC approaches exhibit limitations when confronted with dynamic environmental variations or nonlinear stochastic disturbances. Despite the increasing maturity of ILC theory, control for strongly nonlinear systems remains a core bottleneck : existing ILC methods mostly rely on structural assumptions or Lipschitz conditions, making it difficult to adapt to complex nonlinear scenarios without prior information and lacking a universal design framework [16].

With the rapid advancement of machine learning and artificial intelligence, particularly reinforcement learning (RL), data-driven adaptive control methods have garnered significant attention in batch process optimization [4, 17-24]. RL dynamically optimizes control policies through trial-and-error interaction with the environment, demonstrating robustness in unknown or difficult-to-model systems. However, RL training typically requires extensive random exploration, particularly during early stages, which may compromise system stability or pose safety hazards. In process industries, where production environments are inherently complex and potentially hazardous, reliance solely on RL exploration mechanisms could result in unacceptable control errors and safety concerns. Consequently, developing RL-based batch process control policies that maintain safety guarantees represents a critical research challenge. Bloor et al. [25] recently proposed a Control-Informed Reinforcement Learning (CIRL) method for chemical processes that integrates the strengths of PID control and deep reinforcement learning (DRL) to enhance control performance, robustness, and sample efficiency. This approach represents a customized framework that embeds specific controller structures to guide the RL training process. However, their method depends on predetermined neural network architectures (e.g., PID-style neural networks) as the controller structure. While this design is well-suited for scenarios with known controllers, it presents challenges for generalization to broader control applications. In batch processes characterized by significant noise interference, this framework may inadequately address stringent control accuracy and stability requirements, as it lacks the capability to dynamically adapt the controller's internal structure to mitigate the amplification effects of noise on derivative operations.

To address the dual challenges of safety risks arising from random exploration in RL agents and the extensive real-world iterations required for convergence, this paper proposes a control-informed learning and adaptation approach that integrates ILC with RL, termed Iterative Learning Control-Informed Reinforcement Learning (IL-CIRL). IL-CIRL leverages domain knowledge of batch process dynamics to guide the RL-based controller's learning process under multiple complex disturbances, including both periodic and non-periodic disturbances as well as continuous dynamic mode transitions in batch processes. By synthesizing ILC principles with RL [26], IL-CIRL transforms RL from a purely exploration-dependent black-box process into a structured learning framework that progressively enhances model accuracy and control stability. This approach enables intelligent optimization control of batch processes while maintaining safety and stability constraints, thereby establishing a novel paradigm for continuous adaptive optimization control. Specifically, IL-CIRL employs a hierarchical ILC informer based on Kalman filter [26, 27] state estimation within the iterative learning framework [10, 28] to derive a control law that satisfies system constraints and ensures stability. This control law subsequently guides the learning process of the RL agent. Within this hierarchical ILC architecture, the Kalman filter serves a dual purpose: it provides real-time estimation of process disturbances while simultaneously delivering accurate system state feedback to the RL agent. This dual functionality mitigates safety risks associated with random exploration and enables the agent to accurately capture process states in complex environments characterized by dynamic behavior, nonlinearity, and varying operating conditions. Furthermore, the ILC-based two-layer hierarchical control strategy systematically guides the RL agent toward asymptotic convergence while enabling progressive optimization of the control policy.

This paper makes the following key contributions:

1. **Integration of ILC and RL for batch process optimization.** The proposed IL-CIRL framework represents a pioneering effort to incorporate control information into RL training processes. This integration provides a novel approach to batch process optimization that avoids unsafe exploration while guaranteeing convergence.
2. **Real-time state estimation using Kalman filtering.** A Kalman filter-based state estimation method is developed to enable the RL agent to accurately capture process states in dynamic and nonlinear environments, thereby facilitating optimized control strategies.
3. **Enhanced robustness through ILC-informed reinforcement learning.** By integrating ILC-guided strategies with RL, various disturbances and inherent system nonlinearities in batch processes are effectively addressed, leading to significant improvements in overall optimization and control performance.
4. **Safe deployment strategy for real-world implementation.** Safety in practical batch process control is ensured through offline pre-training with a hierarchical ILC informer and online safe implementation based on a weighted fusion strategy. This approach enables a secure transition to RL-based end-to-end control.

The remainder of this paper are structured as follows: Section 2 presents the background, challenges, and problem definition of batch process control, while analyzing and describing the disturbance characteristics of batch processes; Section 3 details the framework design of the IL-CIRL algorithm, including the batch-to-batch and within-batch ILC control laws based on Kalman filters and the IL-CIRL algorithm process; Section 4 provides the experimental design and results analysis; Section 5 summarizes the key contributions of this study and proposes future research directions.

## 2 Problem statement

### 2.1 Overview of batch process models

Batch processes differ fundamentally from continuous chemical processes in that they exhibit distinct multi-phase or multi-stage transient characteristics. At the batch level, these processes display time-varying nonlinear behavior that varies with operating conditions. Consequently, conventional state-space models based on linear time-invariant (LTI) systems are inadequate for capturing the dynamic modal behavior of batch processes [29]. This study employs a linear time-varying (LTV) model to characterize the process dynamics of a batch process operating along its entire optimal nominal trajectory. The model is expressed as follows:

$$\begin{aligned} \mathbf{x}_k(t+1) &= \mathbf{A}(t)\mathbf{x}_k(t) + \mathbf{B}_u(t)\mathbf{u}_k(t) + \mathbf{B}_d(t)\mathbf{d}_k(t) \\ \mathbf{z}_k(t) &= \mathbf{F}(t)\mathbf{x}_k(t) + \mathbf{m}_k(t) \\ \mathbf{y}_k(t) &= \mathbf{C}(t)\mathbf{x}_k(t) + \mathbf{n}_k(t) \end{aligned} \quad (1)$$

where $k$ denotes the batch index, and $t \in \mathbf{I}_{0:T}$ is the time index within the $k$-th batch, $\mathbf{x}_k(t) \in \mathbb{R}^{n_x}$, $\mathbf{u}_k(t) \in \mathbb{R}^{n_u}$, $\mathbf{d}_k(t) \in \mathbb{R}^{n_d}$, $\mathbf{z}_k(t) \in \mathbb{R}^{n_z}$, and $\mathbf{z}_k(t) \in \mathbb{R}^{n_z}$ respectively represent the system state, control action, process disturbance, observation signal, and product quality at time, $\mathbf{A}(t)$, $\mathbf{B}_u(t)$, $\mathbf{B}_d(t)$, $\mathbf{F}(t)$, and $\mathbf{C}(t)$ are the time-varying dynamic matrices in the LTV state-space model, $\mathbf{m}_k(t) \in \mathbb{R}^{n_z}$ and $\mathbf{n}_k(T) \in \mathbb{R}^{n_y}$ are the measurement noises of the observation variable and quality variable, respectively, assumed to be white noises, i.e., $\mathbf{m}_k(t) \sim N(\mathbf{0}, \sigma_m^2)$ and $\mathbf{n}_k(T) \sim N(\mathbf{0}, \sigma_n^2)$.

Product quality in batch processes is typically not measured in real-time. Instead, quality measurements are obtained only after batch completion through laboratory analysis. Consequently, terminal product quality, $\mathbf{y}_k(T)$, is the more relevant metric for batch process optimization and control. The corresponding time-varying matrix is denoted as $\mathbf{C}(T)$.

Given that Eq.(1) still presents the system in a time-scaled state-space form, it is necessary to reformulate the model into a more compact representation suitable for batch-oriented optimization and control. Specifically, the state-space model can be expressed as follows:

$$\begin{aligned}
\mathbf{x}_k &= \Phi \mathbf{x}_k(0) + \Psi_u \mathbf{u}_k + \Psi_d \mathbf{d}_k \\
\mathbf{z}_k &= \Omega \mathbf{x}_k + \mathbf{m}_k \\
\mathbf{y}_k(T) &= \Gamma \mathbf{x}_k + \mathbf{n}_k(T)
\end{aligned} \qquad (2)$$

where the system states, inputs, disturbances, observations, and measurement noises are reformulated as one-dimensional vectors representing the entire batch, with their specific definitions given as follows:

$$\begin{aligned}
\mathbf{x}_k &= \begin{bmatrix} \mathbf{x}_k(1)^T & \mathbf{x}_k(2)^T & \cdots & \mathbf{x}_k(T)^T \end{bmatrix}^T, \\
\mathbf{u}_k &= \begin{bmatrix} \mathbf{u}_k(0)^T & \mathbf{u}_k(1)^T & \cdots & \mathbf{u}_k(T-1)^T \end{bmatrix}^T, \\
\mathbf{d}_k &= \begin{bmatrix} \mathbf{d}_k(0)^T & \mathbf{d}_k(1)^T & \cdots & \mathbf{d}_k(T-1)^T \end{bmatrix}^T, \\
\mathbf{z}_k &= \begin{bmatrix} \mathbf{z}_k(1)^T & \mathbf{z}_k(2)^T & \cdots & \mathbf{z}_k(T)^T \end{bmatrix}^T, \\
\mathbf{m}_k &= \begin{bmatrix} \mathbf{m}_k(1)^T & \mathbf{m}_k(2)^T & \cdots & \mathbf{m}_k(T)^T \end{bmatrix}^T
\end{aligned} \qquad (3)$$

The Hankel matrices for the system state, system input, and system output, as defined in Eq.(3), are calculated as follows:

$$\begin{aligned}
\Phi &= \begin{bmatrix} \mathbf{A}(0) \\ \mathbf{A}(1)\mathbf{A}(0) \\ \vdots \\ \prod_{i=0}^{T-1} \mathbf{A}(i) \end{bmatrix}, \\
\Omega &= \mathrm{diag}\left(\begin{bmatrix} \mathbf{F}(1) & \mathbf{F}(2) & \cdots & \mathbf{F}(T) \end{bmatrix}\right), \\
\Gamma &= \begin{bmatrix} \mathbf{0} & \cdots & \mathbf{C}(T) \end{bmatrix}
\end{aligned} \qquad (4)$$

$$\Psi_u = \begin{bmatrix}
\mathbf{B}_u(0) & 0 & \cdots & 0 \\
\mathbf{A}(1)\mathbf{B}_u(0) & \mathbf{B}_u(1) & \cdots & 0 \\
\vdots & \vdots & \ddots & \vdots \\
\prod_{i=0}^{T-1} \mathbf{A}(i)\mathbf{B}_u(0) & \prod_{i=1}^{T-1} \mathbf{A}(i)\mathbf{B}_u(1) & \cdots & \mathbf{B}_u(T-1)
\end{bmatrix} \qquad (5)$$
$$= \begin{bmatrix} \Psi_u(0) & \Psi_u(1) & \cdots & \Psi_u(T-1) \end{bmatrix}$$

$$\Psi_d = \begin{bmatrix}
B_d(0) & 0 & \cdots & 0 \\
A(1)B_d(0) & B_d(1) & \cdots & 0 \\
\vdots & \vdots & \ddots & \vdots \\
\prod_{i=1}^{T-1} A(i)B_d(0) & \prod_{i=2}^{T-1} A(i)B_d(1) & \cdots & B_d(T-1)
\end{bmatrix} \qquad (6)$$
$$= \begin{bmatrix} \Psi_d(0) & \Psi_d(1) & \cdots & \Psi_d(T-1) \end{bmatrix}$$

Typically, the economic optimization objectives for batch processes during operation can be divided into two components: stage cost and terminal cost, which are calculated as follows:

$$V(\mathbf{x},\mathbf{u}) = \sum_{t=0}^{T-1} l(\mathbf{x}(t),\mathbf{u}(t)) + Q(\mathbf{y}(T)) \tag{7}$$

where $l: \mathbb{R}^{n_x} \times \mathbb{R}^{n_u} \to \mathbb{R}$ represents the operational cost incurred at each time instant before the completion of the batch, and $Q: \mathbb{R}^{n_y} \to \mathbb{R}$ denotes the economic cost associated with the terminal product quality at the end of the batch. According to the hierarchical control architecture commonly adopted in process industries, the optimization control problem for batch processes is typically addressed in multiple layers. The upper-layer Real-Time Optimization (RTO) module solves the economic optimization problem while considering constraints such as system dynamics and parameter bounds. The resulting nominal optimal trajectory, $(\mathbf{x}_{\text{nom}}, \mathbf{u}_{\text{nom}})$, is then provided as a reference trajectory to lower-layer controllers such as Model Predictive Control (MPC) and Proportional-Integral-Derivative (PID) schemes. In practical implementation, the LTV state-space model described in Eq.(1) is linearized along the reference trajectory. By using the batch-wise state-space model in Eq.(2) and the economic objective function in Eq.(7), the optimal control law is continuously computed and updated for each batch in accordance with the predefined economic objectives.

## 2.2 Analysis and description of disturbance characteristics of batch processes

Traditional batch process control schemes, such as ILC, are typically designed for scenarios characterized by fixed inter-batch disturbances. These methods utilize disturbance information from previous batches to iteratively optimize control inputs for subsequent batches. However, when addressing random uncertainties within a batch, it is essential to consider disturbance characteristics along the temporal axis. Specifically, batch process disturbances can be classified according to their properties: deterministic repetitive disturbances, random non-repetitive disturbances, and random variations of deterministic disturbances across adjacent batches. Given that process disturbances comprise both deterministic/repetitive and random/non-repetitive components, analyzing the correlation between disturbances and process dynamics is necessary.

It is assumed that the disturbances in a batch process consist of a superposition of deterministic and random disturbances, described as follows:

$$\mathbf{d}_k = \bar{\mathbf{d}}_k + \mathbf{v}_k \tag{8}$$

where $\mathbf{d}_k = \begin{bmatrix} \mathbf{d}_k(0)^T & \mathbf{d}_k(1)^T & \ldots & \mathbf{d}_k(T-1)^T \end{bmatrix}^T$ represents the disturbance vector composed of process disturbances at all time instants of the $k$-th batch, $\bar{\mathbf{d}}_k = \begin{bmatrix} \bar{\mathbf{d}}_k(0)^T & \bar{\mathbf{d}}_k(1)^T & \ldots & \bar{\mathbf{d}}_k(T-1)^T \end{bmatrix}^T$ denotes the unknown deterministic/repetitive component of the process disturbance, and $\mathbf{v}_k = \begin{bmatrix} \mathbf{v}_k(0)^T & \mathbf{v}_k(1)^T & \ldots & \mathbf{v}_k(T-1)^T \end{bmatrix}^T$ corresponds to the random or non-repetitive component. The deterministic disturbances across batches may exhibit slight random variations, and the disturbances in adjacent batches may also change randomly. This relationship can be described as follows:

$$\bar{\mathbf{d}}_{k+1} = \bar{\mathbf{d}}_k + \mathbf{w}_k \tag{9}$$

where $\bar{\mathbf{d}}_{k+1}$ represents the deterministic disturbance of the next batch, and $\mathbf{w}_k = \begin{bmatrix} \mathbf{w}_k(0)^T & \mathbf{w}_k(1)^T & \ldots & \mathbf{w}_k(T-1)^T \end{bmatrix}^T$ denotes the random variation in the deterministic process disturbance between adjacent batches. Both disturbances $\mathbf{w}_k(t)$, $t \in \mathbf{I}_{0:T-1}$ and $\mathbf{v}_k(t)$, $t \in \mathbf{I}_{0:T-1}$ in Eqs.(8) and (9) are assumed to be white noises, with distributions denoted by $\mathbf{v}_k(t) \sim N(\mathbf{0}, \sigma_v^2)$ and $\mathbf{w}_k(t) \sim N(\mathbf{0}, \sigma_w^2)$, respectively.

## 2.3 Reconstructing the batch-to-batch state-space model in the batch direction

Modeling the state-space formulation across batches is relatively straightforward. It simply requires substituting the disturbance terms from Eq.(8) into the batch-wise state-space model given in Eq.(2). Accordingly, the state-space representation for batch $k$ can then be rewritten as follows:

$$\mathbf{x}_k = \Phi \mathbf{x}_k(0) + \Psi_u \mathbf{u}_k + \Psi_d (\bar{\mathbf{d}}_k + \mathbf{v}_k) \tag{10}$$

Similarly, the state-space representation for batch $k+1$ can be expressed as:

$$\mathbf{x}_{k+1} = \Phi \mathbf{x}_{k+1}(0) + \Psi_u \mathbf{u}_{k+1} + \Psi_d (\bar{\mathbf{d}}_{k+1} + \mathbf{v}_{k+1}) \tag{11}$$

By applying a difference operation to the above state-space models, the incremental state-space representations, the incremental state-space model can be formulated as:

$$\mathbf{x}_{k+1} = \mathbf{x}_k + \Psi_u \Delta \mathbf{u}_{k+1} + \Phi(\mathbf{x}_{k+1}(0) - \mathbf{x}_k(0)) + \Psi_d (\mathbf{w}_k + \mathbf{v}_{k+1} - \mathbf{v}_k) \tag{12}$$

Assuming that the initial state fluctuations $\mathbf{x}_{k+1}(0) - \mathbf{x}_k(0)$ between adjacent batches are random variables, their effect on the state diminishes over time. Consequently, the third term in Eq.(12) can be neglected, leading to the following simplified expression:

$$\mathbf{x}_{k+1} = \mathbf{x}_k + \Psi_u \Delta \mathbf{u}_{k+1} + \Psi_d (\mathbf{w}_k + \mathbf{v}_{k+1} - \mathbf{v}_k) \tag{13}$$

This incremental state-space model does not explicitly include the process disturbance $\mathbf{d}_k$

from Eq.(2); instead, it is reformulated to incorporate the corresponding random variable $\mathbf{w}_{k-1} + \mathbf{v}_k - \mathbf{v}_{k-1}$. Accordingly, the state-space representation for batch $k$ can be rewritten as:

$$\begin{aligned}
\mathbf{x}_k &= \mathbf{x}_{k-1} + \Psi_u \Delta \mathbf{u}_k + \Psi_d (\mathbf{w}_{k-1} + \mathbf{v}_k - \mathbf{v}_{k-1}) \\
\mathbf{z}_k &= \Omega \mathbf{x}_k + \mathbf{m}_k \\
\mathbf{y}_k(T) &= \mathbf{C}(T)\mathbf{x}_k + \mathbf{n}_k(T)
\end{aligned} \tag{14}$$

## 2.4 Reconstructing the within-batch state-space model in the time direction

Since batch processes exhibit dynamic characteristics in both the batch and time directions, relying solely on the batch-level state-space model described in Eq.(14) may fail to account for transient disturbances occurring within a batch. This oversight could result in control errors that compromise the optimal control performance of subsequent batches. Therefore, it is essential to develop a within-batch state-space model that captures the dynamics along the time direction.

To this end, Eq.(14) must be reformulated in a form suitable for within-batch ILC, explicitly representing the influence of control signals at each time step on the system state. The state-space model can then be expressed as:

$$\mathbf{x}_k = \mathbf{x}_{k-1} + \sum_{i=0}^{T-1} \Psi_u(i) \Delta \mathbf{u}_k(i) + \sum_{i=0}^{T-1} \Psi_d(i)(\mathbf{w}_{k-1}(i) + \mathbf{v}_k(i) - \mathbf{v}_{k-1}(i)) \tag{15}$$

where $i$ denotes the time instant within a batch, and the incremental state change between adjacent batches is the superposition of contributions from all time instants $i = 0, 1, \cdots T$ to the system dynamics. If only the contributions of the control inputs up to the current time instant $t$ are considered, i.e., $\Delta \mathbf{u}_k(t) = \Delta \mathbf{u}_k(t+1) = \cdots = \Delta \mathbf{u}_k(T-1) = \mathbf{0}$, the cumulative state response to time $t$ can be defined as $\mathbf{x}_{h;k}(t)$, which is calculated as follows:

$$\mathbf{x}_{h;k}(t) = \mathbf{x}_{k-1} + \sum_{i=0}^{t-1} \Psi_u(i) \Delta \mathbf{u}_k(i) + \sum_{i=0}^{t-1} \Psi_d(i)(\mathbf{w}_{k-1}(i) + \mathbf{v}_k(i) - \mathbf{v}_{k-1}(i)) \tag{16}$$

Similarly, the state of the entire batch at time $t+1$ can be written as:

$$\mathbf{x}_{h;k}(t+1) = \mathbf{x}_{k-1} + \sum_{i=0}^{t} \Psi_u(i) \Delta \mathbf{u}_k(i) + \sum_{i=0}^{t} \Psi_d(i)(\mathbf{w}_{k-1}(i) + \mathbf{v}_k(i) - \mathbf{v}_{k-1}(i)) \tag{17}$$

By taking the difference between Eqs.(16) and (17), the original batch-wise state-space model can be reformulated as an incremental state-space representation along the time axis, given by:

$$\mathbf{x}_{h;k}(t+1) = \mathbf{x}_{h;k}(t) + \Psi_u(t) \Delta \mathbf{u}_k(t) + \Psi_d(t)(\mathbf{w}_{k-1}(t) + \mathbf{v}_k(t) - \mathbf{v}_{k-1}(t)) \tag{18}$$

Corresponding to the incremental state description of $\mathbf{x}_{h;k}(t)$ above, the incremental form of the control trajectory for the entire batch at time $t$ is defined as:

$$\mathbf{u}_{h;k}(t) = \mathbf{u}_{k-1} + \left[ \underbrace{\Delta \mathbf{u}_k(0)^T \ \cdots \ \Delta \mathbf{u}_k(t-1)^T}_{t} \ \underbrace{\mathbf{0}^T \ \cdots \ \mathbf{0}^T}_{T-t} \right]^T \quad (19)$$

Accordingly, the incremental forms of the control trajectory at time $t+1$ and at the terminal time $T$ can be written as:

$$\mathbf{u}_{h;k}(t+1) = \mathbf{u}_{h;k}(t) + \left[ \underbrace{\mathbf{0}^T \ \cdots \ \mathbf{0}^T}_{t} \ \Delta \mathbf{u}_k(t)^T \ \underbrace{\mathbf{0}^T \ \cdots \ \mathbf{0}^T}_{T-t-1} \right]^T \quad (20)$$

$$\begin{aligned}\mathbf{u}_{h;k}(T) &= \mathbf{u}_{h;k}(t) + \left[ \underbrace{\mathbf{0} \ \cdots \ \mathbf{0}}_{t} \ \Delta \mathbf{u}_k(t) \ \underbrace{\Delta \mathbf{u}_k(t+1) \ \cdots \ \Delta \mathbf{u}_k(T-1)}_{T-t-1} \right]^T \\ &= \mathbf{u}_{h;k}(t) + \Delta \mathbf{u}_{h;k}(t) \end{aligned} \quad (21)$$

where $\Delta \mathbf{u}_{h;k}(t)$ is defined as $\Delta \mathbf{u}_{h;k}(t) = [\underbrace{\mathbf{0}^T \ \ldots \ \mathbf{0}^T}_{t}, \underbrace{\Delta \mathbf{u}_k(t)^T \ \Delta \mathbf{u}_k(t+1)^T \ \ldots \ \Delta \mathbf{u}_k(T-1)^T}_{T-t}]^T$. Consequently, the state prediction from the current time $t$ to the terminal time $T$ can be calculated as follows:

$$\begin{aligned}\mathbf{x}_{h;k}(T) &= \mathbf{x}_{h;k}(t) + \sum_{i=t}^{T-1} \Psi_u(i) \Delta \mathbf{u}_k(i) \\ &= \mathbf{x}_{h;k}(t) + \Psi_u \Delta \mathbf{u}_{h;k}(t) \end{aligned} \quad (22)$$

## 3 Methodology

### 3.1 Framework for control-informed reinforcement learning in batch processes via iterative learning control

To address the practical challenges associated with RL-based controllers, including safety hazards arising from random exploration during interaction with industrial plants, and the need for extensive iterations before convergence, this study proposes an iterative learning control-informed reinforcement learning (IL-CIRL) framework. The goal of IL-CIRL is to provide an autonomous learning paradigm that ensures safety and asymptotic convergence for batch process optimization control.

The IL-CIRL framework incorporates a learning guidance mechanism based on Kalman filter-enhanced ILC. The ILC generates constraint-compliant control outputs with provable convergence guarantees, providing robust supervision signals for the RL agent. During the pre-training process, the Kalman filter serves as a dual function: it delivers progressively refined state estimates of process disturbances to the RL agent while simultaneously providing an optimized dynamic model, refined both batch-to-batch and within-batch, to the hierarchical ILC informer. This architecture enables the coordinated iteration between high-level trajectory optimization and low-level control synthesis, maintaining performance across diverse disturbance categories, including repetitive and non-repetitive uncertainties, as well as deterministic/stochastic variations.

The overall architecture of the IL-CIRL approach is illustrated in Fig. 1. The IL-CIRL framework comprises two primary components: a hierarchical ILC informer and a DRL agent. The hierarchical ILC informer employs a cascade control structure to decompose the batch process control problem into two dimensions: within-batch control (temporal axis) and batch-to-batch control (batch axis). Both sub-ILC informers utilize Kalman filtering to mitigate process disturbances originating from multiple sources. The DRL agent is algorithm-agnostic and can be implemented using various RL algorithms; in this study, proximal policy optimization (PPO) is utilized.

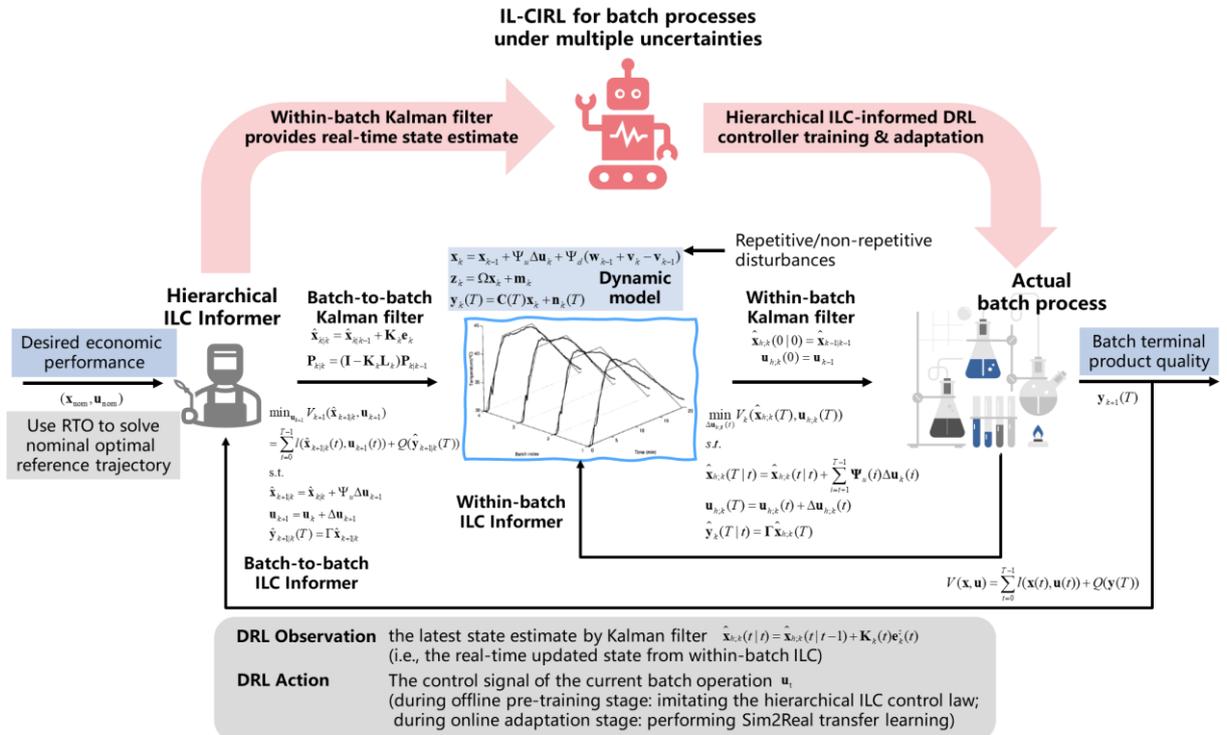

Fig. 1 Overall framework of the proposed IL-CIRL algorithm.

It is important to note that IL-CIRL employs distinct objectives and structural configurations for the offline pre-training and online fine-tuning phases. During offline pre-training, the framework emphasizes control-informed policy learning, wherein the hierarchical ILC informer, constructed using a LTV dynamic model, provides reference control trajectories. At this stage, the DRL agent's primary objective is to replicate the control actions generated by the informer. Conversely, the online training phase prioritizes safety-guaranteed control-informed adaptation. In this phase, the DRL agent and hierarchical ILC informer collaboratively generate hybrid control actions, enabling the DRL agent to progressively refine its imitation-based policy. As training advances, the weighting assigned to the informer decreases systematically, and control authority gradually transfers to the DRL agent until autonomous operation is achieved.

The IL-CIRL framework incorporates ILC and disturbance information into the controller training procedure to guide the RL agent toward safe and progressively convergent control behaviors during both batch-to-batch (episode-level) and within-batch (time-step-level) exploration. The key features of IL-CIRL are as follows:

1. **Integration of ILC structure.** The framework integrates an ILC structure into the training process, enabling control-informed reinforcement learning for batch process optimization.
2. **Dual-layer ILC controller design.** A Kalman filter-based dual-layer ILC controller operates at both the batch-to-batch and within-batch levels, guaranteeing progressive convergence and providing safety-assured guidance for offline pre-training and online execution.
3. **Disturbance handling.** The framework addresses disturbances of varying characteristics (repetitive/non-repetitive, deterministic/stochastic) by leveraging Kalman filtering to mitigate model-plant mismatch in ILC, thereby transforming the uncertainty-induced perturbation problem into a state estimation problem.
4. **Offline pre-training capability.** During offline pre-training, the within-batch Kalman filter provides real-time state estimation to the RL agent, eliminating the need for direct interaction with the actual industrial process.
5. **Adaptive online execution strategy.** During online execution, an adaptive weighting strategy is employed whereby the control signal is initially generated jointly by the hierarchical ILC and RL agent. The weight gradually shifts toward the RL component as training stabilizes, ultimately yielding a fully RL-based end-to-end controller without compromising safety.

**3.2 Control-informed policy learning during offline pre-training processes**

**3.2.1 The proposed offline pre-training scheme**

A primary challenge in applying RL to industrial control systems is its dependence on stochastic exploration during training, which can result in unsafe or unpredictable control actions. In batch process optimization, particularly during early training stages, such behavior poses significant risks, including process instability and unacceptable operational hazards. To mitigate these concerns, the IL-CIRL framework implements a control-informed policy learning strategy during offline pre-training. This approach leverages the complementary strengths of ILC and RL to establish a safe learning paradigm that eliminates hazardous exploration prior to deployment in real-world processes.

During offline pre-training, the IL-CIRL agent operates independently of the actual industrial system. Instead, it learns a preliminary policy within a simulated environment, where ILC-based control information serves as a critical bridge to the previously obtained LTV dynamic model. Specifically, the Kalman filter-based control-informed learning mechanism provides the RL agent with a stable, asymptotically convergent dynamic model for agent-environment interactions. The Kalman filter continuously estimates process disturbances and generates state feedback to align the RL agent's actions with those of the hierarchical ILC controller, thereby guiding the pre-trained RL agent's action output toward optimal performance. This mechanism effectively integrates the convergence and stability guarantees of Kalman filter-based ILC into the RL training process. Fig. 2 illustrates how hierarchical ILC functions as an expert controller, providing control information to guide the IL-CIRL agent during offline pre-training.

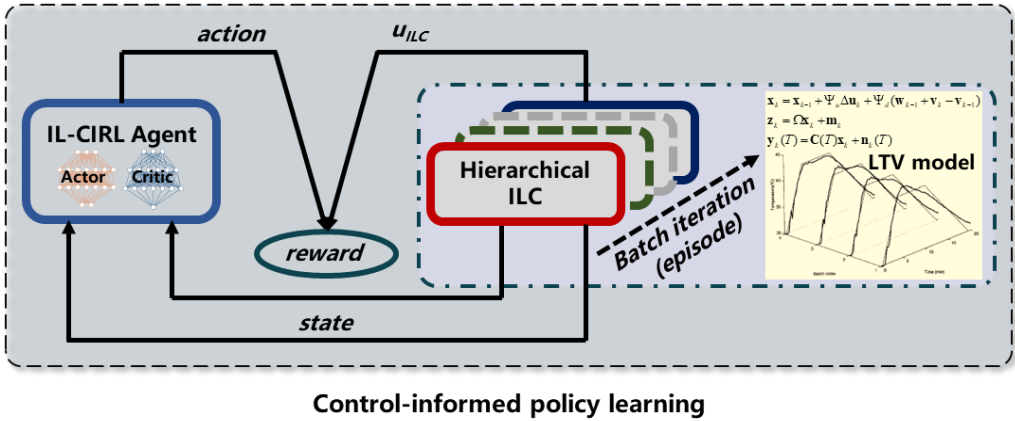

Fig. 2 Schematic diagram of control-informed policy learning during offline pre-training processes

The primary objective of this control-informed guidance strategy is to mitigate exploration-related risks by providing accurate system dynamics information, thereby establishing a stable pre-training environment for policy optimization. Additionally, the Kalman filter enhances training reliability by iteratively refining the system's predictive model to deliver precise state estimates for RL. This approach prevents unstable behaviors arising from model-plant mismatch or process disturbances. Consequently, control information enables the RL agent to progressively adapt to actual process dynamics while steering the policy toward global optimality, ultimately providing robust initialization for subsequent online adaptation.

In summary, the control-informed policy learning strategy during offline pre-training enhances RL training through two key mechanisms: (1) precise system state estimation via Kalman filtering, which enables accurate environment modeling, and (2) elimination of unsafe exploration through informed supervision, which ensures both system safety and learning efficiency during the pre-training phase. The pseudocode of the CIRL algorithm during offline training is shown in Algorithm 1. The detailed design schemes are explained in the following subsections.

```
Algorithm 1 Offline IL-CIRL
Require: actor: Policy network architecture, critic: Value network architecture, agentOpts: PPO agent parameters,
    Hierarchical ILC Informer: ILC algorithm, env: Environment composed of LTV model and hierarchical ILC, T_step = 40:
    Number of time steps per batch
Ensure: Trained PPO agent agent
 1: agent ← PPOAgent(actor, critic, agentOpts)
 2: for k = 1 to N_episodes do
 3:     Reset environment: s_0 ← env.reset()
 4:     Initialize batch trajectory data T ← ∅
 5:     for t = 0 to T_step − 1 do                                          ▷ 40 time steps per batch
 6:         Iterative learning Control generates action: u_ILC ← ILC.computeAction(s_t)
 7:         Environment executes action: s_{t+1}, r_env ← env.step(u_ILC)
 8:         Agent observes current state: agent.observe(s_t)
 9:         Agent generates action: a_t ← agent.getAction(s_t)              ▷ Used only for reward calculation
10:         Calculate agent reward:
                            r_t = −‖u_ILC − a_t‖^2
11:         Store state-action-reward tuple:
                            T ← T ∪ {(s_t, a_t, r_t, s_{t+1})}
12:         Update state and last action:
                            s_t ← s_{t+1},    a_{t−1} ← a_t
13:     end for
14:     Agent updates policy using current batch data:
                            agent.update(T)
15: end for
16: return agent
```

### 3.2.2 IL-CIRL agent design (offline pre-training process)

This section presents the conceptual design principles of the IL-CIRL agent; detailed network configurations and parameter settings are provided in Section 4.

**1) Conceptual overview of the IL-CIRL pre-training procedure**

The IL-CIRL framework employs a distinctive pre-training approach that differs fundamentally from conventional DRL methodologies. Traditional DRL pre-training involves agent interaction with a simulated environment prior to transfer or fine-tuning for deployment in real systems [30]. In contrast, the IL-CIRL pre-training phase operates without any interaction with the physical plant or process simulator. Instead, the agent interacts exclusively with the hierarchical ILC informer across both the batch and time axes, which functions on the LTV state-space model of batch processes derived in Section 2.

The hierarchical ILC informer incorporates Kalman filtering to address the various disturbances described in Section 2.2, transforming the disturbance-compensated control design problem into a state estimation framework. This approach eliminates the need for the IL-CIRL agent to directly incorporate disturbance terms as RL states. The informer comprises two hierarchical components: a batch-to-batch Kalman filter-based ILC informer and a within-batch Kalman filter-based ILC informer. The former functions as the outer loop in a cascade architecture, transmitting information, including end-of-batch state estimates, to the inner-loop within-batch ILC informer. This hierarchical design enables the framework to simultaneously manage both deterministic persistent (batch-wise) and stochastic transient (time-wise) disturbances. Additionally, the Kalman filter supports the informer's prediction model while providing disturbance-filtered state estimates to the IL-CIRL agent, thereby facilitating seamless integration between ILC and control-informed deep reinforcement learning.

**2) IL-CIRL agent & proximal policy optimization (PPO)**

The IL-CIRL algorithm employs the PPO algorithm as its DRL agent for batch process optimization control. PPO is a model-free, on-policy, policy-gradient-based DRL algorithm that has gained widespread recognition for its stability and robust empirical performance. The algorithm operates by alternating between data sampling through environmental interactions and optimizing a clipped surrogate objective function via stochastic gradient descent [31]. By constraining large deviations between successive policies, which can impede learning, PPO ensures stable policy updates and mitigates the instability inherent in traditional policy gradient methods. Through the introduction of a surrogate objective function that enables mini-batch updates across multiple training steps, PPO addresses the challenge of step size selection present in conventional policy gradient algorithms.

The PPO-Clip variant constrains the magnitude of policy changes, thereby encouraging the updated policy to remain in proximity to its predecessor. In PPO-Clip, the policy update is governed by the following objective:

$$\theta_{k+1} = \arg\max_\theta \mathbb{E}_{s,a\sim\pi_{\theta_k}} \left[ L(s,a,\theta_k,\theta) \right], \tag{23}$$

where $L$ is defined as the surrogate objective function, which depends on the advantage function $A^{\pi_{\theta_k}}(s,a)$:

$$L(s,a,\theta_k,\theta) = \min\left( \frac{\pi_\theta(a|s)}{\pi_{\theta_k}(a|s)} A^{\pi_{\theta_k}}(s,a),\ g(\epsilon, A^{\pi_{\theta_k}}(s,a)) \right), \tag{24}$$

where

$$g(\epsilon, A) = \begin{cases} (1+\epsilon)A & A \geq 0 \\ (1-\epsilon)A & A < 0. \end{cases} \tag{25}$$

Therefore, the clipping operation serves as a form of regularization that mitigates excessive policy updates. The hyperparameter $\epsilon$ controls the permissible deviation between the old and new policies, ensuring stable and effective learning.

**3) Basic principles of State, Action, and Reward definition**

In control theory and DRL, the terms "state" and "observation" carry distinct meanings. In control systems, state typically refers to internal system variables, such as the system state $\mathbf{x}_k(t) \in \mathbb{R}^{n_x}$ in Eq.(1). In contrast, DRL emphasizes *observations* that satisfy the Markov property. Consequently, within the framework of Markov Decision Processes (MDPs), any variable relevant to the DRL agent's decision-making can be interpreted as state, provided it satisfies the Markov property.

In the proposed IL-CIRL algorithm framework, the DRL state is defined as the state variables of the batch process model. This design choice ensures seamless integration, as the hierarchical ILC informer guides both control-informed offline policy learning and safety-guaranteed online adaptation. This unified definition enables the organic integration of the ILC structure and the DRL agent within a cohesive framework. The DRL *action* corresponds to the control inputs of the batch process, equivalent to the control signal $\mathbf{u}_k(t) \in \mathbb{R}^{n_u}$ in Eq.(1). Specifically, the *state* represents the internal state variables of the batch process, while the *action* represents the manipulated variables.

The reward function settings in IL-CIRL vary between the offline pre-training stage and the online safety adaptation stage. During offline pre-training, IL-CIRL conducts imitation learning based on the hierarchical ILC informer without direct interaction with the actual process. The objective is to replicate expert control patterns under Kalman-filtered feedback, thereby inheriting the expert's convergence and stability characteristics while preventing unsafe exploration.

Consequently, the pre-training reward is designed to quantify the discrepancy between the IL-CIRL agent's actions and those of the hierarchical ILC informer. This approach compels the DRL agent to emulate the hierarchical ILC informer's behavior. Through this control-informed policy learning methodology, expert guidance with guaranteed asymptotic convergence and stability is incorporated into the IL-CIRL pre-training process.

The implementation differs from traditional DRL in two key aspects. First, rather than obtaining state feedback directly from the environment or controlled object, IL-CIRL employs within-batch real-time Kalman filtering for state feedback. Second, the control actions generated by DRL are not transmitted directly to the batch process. The IL-CIRL *state* represents the system state provided by the Kalman filter, and its corresponding *action* should approximate the hierarchical ILC informer's behavior.

Based on these considerations, the pre-training *reward* function is designed as:

$$r_t = -|u_{ILC} - u_{RL}| \tag{26}$$

### 3.2.3 Design of batch-to-batch and within-batch hierarchical ILC informer

During offline pre-training, the hierarchical ILC functions as the informer, governing control across both batch-to-batch and within-batch dimensions. Building upon the LTV state-space predictive models derived in Sections 2.2 and 2.3, the design principles of the Kalman filter-based ILC informer are detailed below.

**1) Batch-to-batch Kalman filter-based ILC for batch process control**

The Kalman filter is employed to iteratively estimate state updates from multiple disturbance sources, based on the batch process state-space model with noise described in Eq.(14). This approach integrates process disturbances into the dynamic model, enabling beneficial disturbances to enhance process economy while mitigating adverse effects. Given the posterior state estimates from the previous batch and the observation variables from the current batch, the Kalman filter estimates the state of the current batch. The state estimation and error covariance matrix for the current batch are calculated through the dynamic model of Eq.(14), with specific formulas provided as follows:

$$\hat{\mathbf{x}}_{k|k-1} = \hat{\mathbf{x}}_{k-1|k-1} + \Psi_u \Delta \mathbf{u}_k \tag{27}$$

$$\begin{aligned}\mathbf{P}_{k|k-1} &= \mathbf{P}_{k-1|k-1} + \mathbf{Q}_k \\ &= \mathbf{P}_{k-1|k-1} + \Psi_d(\mathbf{w}_{k-1} + \mathbf{v}_k - \mathbf{v}_{k-1})(\mathbf{w}_{k-1} + \mathbf{v}_k - \mathbf{v}_{k-1})^T \Psi_d^T \\ &= \mathbf{P}_{k-1|k-1} + \Psi_d \begin{bmatrix} \mathbf{R}_w + 2\mathbf{R}_v & 0 & \cdots & 0 \\ 0 & \mathbf{R}_w + 2\mathbf{R}_v & \cdots & 0 \\ \vdots & \vdots & \ddots & \vdots \\ 0 & 0 & \cdots & \mathbf{R}_w + 2\mathbf{R}_v \end{bmatrix} \Psi_d^T \end{aligned} \quad (28)$$

where $\hat{\mathbf{x}}_{k|k-1}$ represents the state estimation of batch $k$ based on the state observation of batch $k-1$, $\hat{\mathbf{x}}_{k-1|k-1}$ denotes the posterior estimation of the state of batch $k-1$ when the state observation of batch $k-1$ is available, and $\mathbf{P}_{k-1|k-1}$ represents the corresponding state error covariance matrix; $\mathbf{R}_w = \mathbf{w}_k(t)\mathbf{w}_k^T(t)$ and $\mathbf{R}_v = \mathbf{v}_k(t)\mathbf{v}_k^T(t)$. The residual matrix $\mathbf{e}_k$, consisting of the measurement residuals of both the observation variables and the terminal product quality variables, can be computed based on the system dynamics, and its specific expression is given as follows:

$$\mathbf{e}_k = \begin{bmatrix} \mathbf{e}_k^p \\ \mathbf{e}_k^q \end{bmatrix} = \begin{bmatrix} \mathbf{z}_k \\ \mathbf{y}_k(T) \end{bmatrix} - \begin{bmatrix} \Omega \\ \Gamma \end{bmatrix} \mathbf{x}_{k|k-1} = \begin{bmatrix} \mathbf{z}_k \\ \mathbf{y}_k(T) \end{bmatrix} - \mathbf{L}_k \hat{\mathbf{x}}_{k|k-1} \quad (29)$$

Define $\mathbf{L}_k = \begin{bmatrix} \Omega^T & \Gamma^T \end{bmatrix}^T$, then the covariance matrix corresponding to $\mathbf{e}_k$ can be derived from $\mathbf{L}_k$ and $\mathbf{P}_{k|k-1}$. The specific expression is given as follows:

$$\begin{aligned}\mathbf{S}_k &= \mathbf{L}_k \mathbf{P}_{k|k-1} \mathbf{L}_k^T + \mathbf{R}_k = \mathbf{L}_k \mathbf{P}_{k|k-1} \mathbf{L}_k^T + \begin{bmatrix} \mathbf{m}_k \\ \mathbf{n}_k(T) \end{bmatrix}\begin{bmatrix} \mathbf{m}_k \\ \mathbf{n}_k(T) \end{bmatrix}^T \\ &= \mathbf{L}_k \mathbf{P}_{k|k-1} \mathbf{L}_k^T + \begin{bmatrix} \mathbf{R}_m & 0 & \cdots & 0 \\ 0 & \mathbf{R}_m & \cdots & 0 \\ \vdots & \vdots & \ddots & \vdots \\ 0 & 0 & \cdots & \mathbf{R}_n \end{bmatrix}\end{aligned} \quad (30)$$

where $\mathbf{R}_m = \mathbf{m}_k(t)\mathbf{m}_k^T(t)$, and $\mathbf{R}_n = \mathbf{n}_k(t)\mathbf{n}_k^T(t)$. According to the Kalman filter derivation, the optimal Kalman gain $\mathbf{K}_k$ can be calculated as follows:

$$\mathbf{K}_k = \mathbf{P}_{k|k-1} \mathbf{L}_k^T \mathbf{S}_k^{-1} \quad (31)$$

Therefore, the updated state estimation and the corresponding error covariance matrix for the batch process can be expressed in the following iterative update form:

$$\hat{\mathbf{x}}_{k|k} = \hat{\mathbf{x}}_{k|k-1} + \mathbf{K}_k \mathbf{e}_k \quad (32)$$

$$\mathbf{P}_{k|k} = (\mathbf{I} - \mathbf{K}_k \mathbf{L}_k)\mathbf{P}_{k|k-1} \quad (33)$$

Upon obtaining the posterior state estimation for the current batch and its corresponding control sequence, the states of subsequent batches can be continuously predicted using the Kalman filter. In conjunction with the batch process optimization control problem defined in Eq.(7), the Kalman filter-based batch-to-batch ILC control design problem can be formulated as follows:

$$\min_{\mathbf{u}_{k+1}} V_{k+1}(\hat{\mathbf{x}}_{k+1|k}, \mathbf{u}_{k+1}) = \sum_{t=0}^{T-1} l(\hat{\mathbf{x}}_{k+1|k}(t), \mathbf{u}_{k+1}(t)) + Q(\mathbf{y}_{k+1|k}(T))$$
s.t.
$$\hat{\mathbf{x}}_{k+1|k} = \hat{\mathbf{x}}_{k|k} + \Psi_u \Delta \mathbf{u}_{k+1}$$
$$\mathbf{u}_{k+1} = \mathbf{u}_k + \Delta \mathbf{u}_{k+1} \tag{34}$$
$$\hat{\mathbf{y}}_{k+1|k}(T) = \Gamma \hat{\mathbf{x}}_{k+1|k}$$

where the computation of the optimal solution $(\hat{\mathbf{x}}^*_{k+1|k}, \mathbf{u}^*_{k+1})$ and its corresponding economic performance index $V^*_{k+1}$ is achieved through Kalman filtering, thereby eliminating the need for explicit disturbance estimation. The resulting optimal control sequence is applied to the batch process in an open-loop manner, whereas the batch-wise direction follows a closed-loop iterative process based on ILC.

During the batch-to-batch iterative learning process, after the completion of the $k$-th batch, the posterior state estimation $\hat{\mathbf{x}}_{k+1|k+1}$ of batch $k+1$ is continuously updated according to the Kalman filter described in Eqs.(32)-(33), thereby enabling iterative refinement of the ILC-based optimal control solution in Eq.(34).

**2) Within-batch Kalman filter-based ILC for batch process control**

To establish the connection between batch-to-batch and within-batch ILC, the state response at time $t = 0$ in Eq.(18) is defined as the state of the previous batch in the batch-to-batch ILC method; i.e., $\mathbf{x}_{h;k}(0) = \mathbf{x}_{k-1}$. This framework enables integration between the two ILC approaches. Specifically, batch-to-batch ILC functions analogously to the outer loop of cascade control in continuous process control, while within-batch ILC corresponds to the inner loop. The Kalman filter employed in batch-to-batch ILC provides the updated state values at time $t = 0$ for use in within-batch ILC.

Similarly, the Kalman filter used in the batch-to-batch ILC method described above can be applied to state estimation in within-batch ILC. The key distinction lies in the update frequency: batch-to-batch ILC updates the Kalman filter state estimation once per batch, whereas within-batch ILC performs updates at each time step within a batch. For within-batch implementation, the system estimate and error covariance matrix for the entire batch at the current time step can be computed using the dynamic model presented in Eq.(18) as follows:

$$\hat{\mathbf{x}}_{h;k}(t|t-1) = \hat{\mathbf{x}}_{h;k}(t-1|t-1) + \mathbf{\Psi}_u(t-1)\Delta\mathbf{u}_k(t-1) \tag{35}$$

$$\begin{aligned}\mathbf{P}_k(t|t-1) &= \mathbf{P}_k(t-1|t-1) + \mathbf{Q}_k(t) \\ &= \mathbf{P}_k(t-1|t-1) \\ &\quad + \mathbf{\Psi}_d(t)(\mathbf{w}_{k-1}(t) + \mathbf{v}_k(t) - \mathbf{v}_{k-1}(t))(\mathbf{w}_{k-1}(t) + \mathbf{v}_k(t) - \mathbf{v}_{k-1}(t))^T \mathbf{\Psi}_d^T(t) \\ &= \mathbf{P}_k(t-1|t-1) + \mathbf{\Psi}_d(t)(\mathbf{R}_w + 2\mathbf{R}_v)\mathbf{\Psi}_d^T(t)\end{aligned} \tag{36}$$

where $\hat{\mathbf{x}}_{h;k}(t|t-1)$ represents the state estimation at time $t$ based on the state observation at time $t-1$, $\hat{\mathbf{x}}_{h;k}(t-1|t-1)$ denotes the posterior estimation of the state at time $t-1$ when the state observation at time $t-1$ is available, and $\mathbf{P}_k(t-1|t-1)$ is the state error covariance matrix that quantifies the estimation accuracy at time $t-1$; $\mathbf{R}_w = \mathbf{w}_k(t)\mathbf{w}_k^T(t)$ and $\mathbf{R}_v = \mathbf{v}_k(t)\mathbf{v}_k^T(t)$ represent the covariance matrices of the process and measurement noise, respectively. The deviation between the actual measured value and the predicted estimation of the observation variable can then be expressed as follows:

$$\begin{aligned}\mathbf{e}_k^z(t) &= \mathbf{z}_k(t) - \mathbf{F}(t)\mathbf{H}(t)\hat{\mathbf{x}}_k(t|t-1), \\ \mathbf{H}(t) &= \begin{bmatrix} \underbrace{0 \quad 0}_{t-1} & \mathbf{I} & \underbrace{0 \quad 0}_{T-t} \end{bmatrix}\end{aligned} \tag{37}$$

where $\mathbf{e}_k^z(t)$ represents the measurement residual of the observation variable. Define $\mathbf{L}_k(t) = \mathbf{F}(t)\mathbf{H}(t)$, then the covariance matrix of the measurement residual can be calculated as follows:

$$\begin{aligned}\mathbf{S}_k(t) &= \mathbf{L}_k(t)\mathbf{P}_k(t|t-1)\mathbf{L}_k(t)^T + \mathbf{R}_t \\ &= \mathbf{L}_k(t)\mathbf{P}_k(t|t-1)\mathbf{L}_k(t)^T + \mathbf{m}_k(t)\mathbf{m}_k(t)^T \\ &= \mathbf{L}_k(t)\mathbf{P}_k(t|t-1)\mathbf{L}_k(t)^T + \mathbf{R}_m\end{aligned} \tag{38}$$

Based on the Kalman filter derivation, the optimal Kalman gain $\mathbf{K}_k(t)$ can be calculated as follows:

$$\mathbf{K}_k(t) = \mathbf{P}_k(t|t-1)\mathbf{L}_k^T(t)\mathbf{S}_k^{-1}(t) \tag{39}$$

Similarly, the updated state estimation of the batch process and its associated error covariance matrix can be expressed in the following iterative update form:

$$\hat{\mathbf{x}}_{h;k}(t|t) = \hat{\mathbf{x}}_{h;k}(t|t-1) + \mathbf{K}_k(t)\mathbf{e}_k^z(t) \tag{40}$$

$$\mathbf{P}_k(t|t) = (\mathbf{I} - \mathbf{K}_k(t)\mathbf{L}_k(t))\mathbf{P}_k(t|t-1) \tag{41}$$

Once the posterior estimate of the current state and the corresponding control sequence are obtained, the states at subsequent time steps can be continuously predicted using the Kalman filter. It should be noted that since within-batch ILC is implemented on the basis of batch-to-batch ILC, each batch must be initialized at $t=0$ using the Kalman filter result from the previous batch's batch-to-batch ILC, i.e., $\mathbf{P}_k(0|0) = \mathbf{P}_{k-1}$, $\hat{\mathbf{x}}_{h;k}(0|0) = \hat{\mathbf{x}}_{k-1|k-1}$ and $\mathbf{u}_{h;k}(0) = \mathbf{u}_{k-1}$. Combined with the definition of the batch process optimization control problem described in Eq.(7), the within-batch ILC control design problem based on the Kalman filter can be formulated as follows:

$$\min_{\Delta \mathbf{u}_{h;k}(t)} V_k(\hat{\mathbf{x}}_{h;k}(T), \mathbf{u}_{h;k}(T))$$

s.t.

$$\hat{\mathbf{x}}_{h;k}(T|t) = \hat{\mathbf{x}}_{h;k}(t|t) + \sum_{i=t+1}^{T-1} \mathbf{\Psi}_u(i) \Delta \mathbf{u}_k(i) \tag{42}$$
$$\mathbf{u}_{h;k}(T) = \mathbf{u}_{h;k}(t) + \Delta \mathbf{u}_{h;k}(t)$$
$$\hat{\mathbf{y}}_k(T|t) = \mathbf{\Gamma} \hat{\mathbf{x}}_{h;k}(T)$$

where the optimal solution $\Delta \mathbf{u}_{h;k}^*(t)$ represents the optimal control sequence from time $t$ to time $T-1$ in batch $k$, and only $\Delta \mathbf{u}_k^*(t)$ at time $t$ is applied to the batch process during the within-batch closed-loop operation.

During the within-batch iterative learning process, upon completion of time step $t$, the posterior state estimate $\hat{\mathbf{x}}_{h;k}(t+1|t+1)$ at time $t+1$ continues to be updated according to the Kalman filtering of Eqs.(40)-(41), thereby continuous iteration of the ILC optimization control solution given in Eq.(42).

Based on the analysis and derivation presented above, the batch-to-batch ILC employing a Kalman filter addresses repetitive disturbances along the batch axis. Subsequently, within-batch ILC further mitigates non-repetitive disturbances along the time axis. (Note: if disturbances in the time axis exhibit repetitive behavior, they can be incorporated into the batch-to-batch ILC disturbance framework.) This two-tier approach enables within-batch control to be completed before random disturbances propagate through the ILC iterative process of subsequent batches.

### 3.3 Safety-guaranteed control-informed adaptation during online training process

During the online training stage, the RL agent interacts with the actual process and refines its policy based on observed feedback. However, the stochastic nature of DRL exploration, combined with the complex dynamics and inherent uncertainties of industrial systems, can result in initially unstable or unsafe agent actions. To mitigate this risk, this study proposes a safety-guaranteed control-informed adaptation strategy in which the hierarchical ILC informer operates in parallel with the RL agent during online implementation and adaptation, as illustrated in Fig. 3. Both components collaborate to generate the final control signal applied to the batch process. This approach mitigates safety hazards during initial exploration while preserving learning efficiency and convergence. The integration of Kalman filtering facilitates real-time state estimation, enabling the agent to accurately assess process conditions at each training iteration. This real-time state feedback reduces control errors in the IL-CIRL agent arising from model-plant mismatch and ensures progressive policy convergence through precise estimation of system dynamics. The pseudocode for the CIRL algorithm during online adaptation is presented in Algorithm 2.

---

**Algorithm 2** Online IL-CIRL

**Require:** actor: Policy network architecture, critic: Value network architecture, agentOpts: PPO agent parameters, ILC: Iterative Learning Control algorithm, env: Actual batch process, $T_{\text{step}} = 40$: Number of time steps per batch, $K = 1000$: Total number of executions, $\alpha, \beta$: Reward function coefficients, $x_{\text{ref}}$: Reference trajectory

**Ensure:** Trained hybrid control agent

1: agent ← PPOAgent(actor, critic, agentOpts)
2: Initialize last agent action: $a_{-1} \leftarrow 0$
3: **for** $k = 1$ to $N_{\text{episodes}}$ **do**  ▷ Main training loop
4:     Reset environment: $s_0 \leftarrow$ env.reset()
5:     Initialize batch trajectory data $\mathcal{T} \leftarrow \emptyset$
6:     **for** $t = 0$ to $T_{\text{step}} - 1$ **do**  ▷ 40 time steps per batch
7:         ILC generates action: $u_{ILC} \leftarrow$ ILC.computeAction($s_t$)
8:         Agent generates action: $a_t \leftarrow$ agent.getAction($s_t$)
9:         Calculate combined action weight:
$$\theta_t = (1 - \exp(-1.1 \cdot |a_t - u_{ILC}|)) \cdot \frac{K-k}{K}$$
10:        Compute hybrid control action:
$$u = \theta_t \cdot u_{ILC} + (1 - \theta_t) \cdot a_t$$
11:        Environment executes hybrid action:
$$s_{t+1}, x_{est} \leftarrow \text{env.step}(u)$$
12:        Calculate total agent reward:
$$r_{continous} = - \cdot |a_t - u_{ILC}| \quad r_t = \alpha \cdot r_{continous} + \beta \cdot r_{discrete}$$
13:        Store trajectory data:
$$\mathcal{T} \leftarrow \mathcal{T} \cup \{(s_t, a_t, u_{ILC}, u, r_t, s_{t+1}, \theta_t)\}$$
14:        Update state and last action:
$$s_t \leftarrow s_{t+1}, \quad a_{t-1} \leftarrow a_t$$
15:     **end for**
16:     Agent updates policy using buffer data: agent.update($\mathcal{T}$)
17: **end for**
18: **return** agent

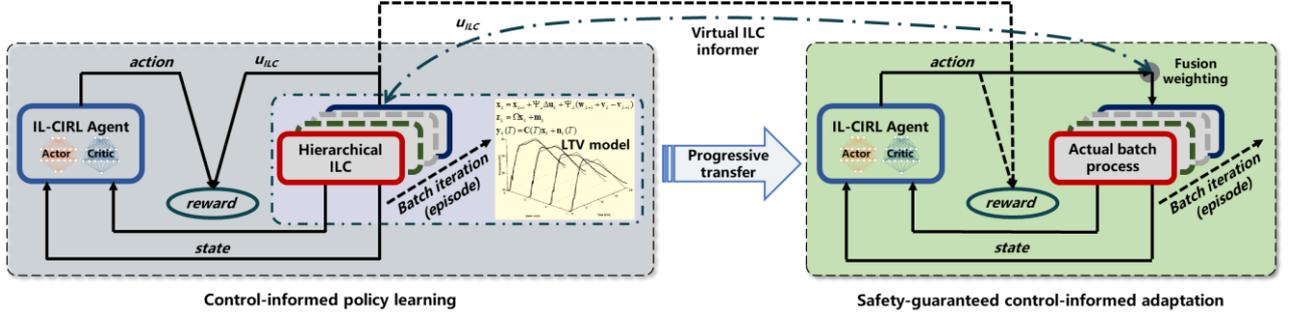

Fig. 3 Schematic diagram of safety-guaranteed control-informed adaptation during online training process

Following the completion of offline pre-training, the IL-CIRL agent is deployed for online adaptation with the actual process. During this phase, both the hierarchical ILC controller and the RL agent generate control signals concurrently to mitigate irrational behaviors. These signals are integrated through an adaptive weighting mechanism, formulated as follows:

$$u = \theta_t \cdot u_{ILC} + (1-\theta_t) \cdot a_t \tag{43}$$

where the weight $\theta$ is used to adjust the control contribution of the DRL policy to the actual batch process. Intuitively, when the policy output of the RL agent deviates significantly from the "expert-level" control signal provided by the ILC informer, the weight of the agent's action should be reduced to mitigate irrational or unsafe behaviors. The weight $\theta$ is updated during online training and adaptation according to following expression:

$$\theta_t = (1 - e^{-(1.1 \cdot |a_t - u_{ILC}|)}) \cdot \frac{K-t}{K} \tag{44}$$

In Eq.(43) the first term captures the relationship between the ILC controller and the RL agent. As the performance gap between them decreases, the corresponding weight decays exponentially. The second term serves as a training-time penalty to prevent the RL agent from relying excessively on the weight and neglecting its own policy updates. By applying this time-dependent penalty after assessing the gap, the agent is encouraged to gradually optimize its policy and ultimately operate independently of the ILC controller's assistance.

Meanwhile, the RL agent adopts the process variables $\hat{x}$ filtered via a Kalman filter as its state representation. Based on this state, the optimization objective and constraints for online training of the agent are formulated as follows:

$$\min_\pi \left( \alpha |u_{ILC} - \pi(\hat{x})| + \beta(\hat{x} - x_{ref}) \right)$$

$$\text{s.t. } \hat{x}_{h_i,k}(T|t) = \hat{x}_{h_i,k}(t|t) + \sum_{i=t+1}^{T-1} \Psi_u(i)\Delta u_k(i) \quad (45)$$

$$u_{h_i,k}(T) = u_{h_i,k}(t) + \Delta u_{h_i,k}(t)$$

$$\hat{y}_k(T|t) = \Gamma \hat{x}_{h_i,k}(T)$$

In other words, the IL-CIRL agent fully leverages the Kalman filtered state in each training iteration or episode, thereby preventing control anomalies due to model adaptation and enabling effective policy optimization.

## 4 Results and Discussion

### 4.1 Experimental setup of batch process optimization control

#### 4.1.1 Batch reaction process description

Table 1 Parameter setting of batch process chemical reactor

| Parameters | Values | Parameters | Values |
|---|---|---|---|
| $\alpha_1$ | $4000 \text{L} \cdot \text{mol}^{-1} \cdot \text{s}^{-1}$ | $\lambda_1$ | $-1.8 \times 10^5 \text{cal} \cdot \text{mol}^{-1}$ |
| $\alpha_2$ | $6.2 \times 10^5 \text{s}^{-1}$ | $\lambda_2$ | $-2.25 \times 10^5 \text{cal} \cdot \text{mol}^{-1}$ |
| $E_1$ | $5000 \text{cal} \cdot \text{g}^{-1} \cdot \text{mol}^{-1}$ | $C_P$ | $1000 \text{cal} \cdot \text{kg}^{-1} \cdot \text{K}^{-1}$ |
| $E_2$ | $10000 \text{cal} \cdot \text{g}^{-1} \cdot \text{mol}^{-1}$ | $C_J$ | $1000 \text{cal} \cdot \text{kg}^{-1} \cdot \text{K}^{-1}$ |
| $R$ | $2 \text{cal} \cdot \text{mol}^{-1} \cdot \text{K}^{-1}$ | $\rho$ | $0.8 \text{kg} \cdot \text{L}^{-1}$ |
| $V$ | $1200 \text{L}$ | $\rho_J$ | $0.8 \text{kg} \cdot \text{L}^{-1}$ |
| $V_J$ | $1200 \text{L}$ | $A_o$ | $525 \text{dm}^2$ |
| $h_{ow}$ | $10850 \text{cal} \cdot \text{min}^{-1} \cdot \text{K}^{-1} \cdot \text{dm}^{-2}$ | | |

In this section, a typical batch reaction process is used to verify the effectiveness of the proposed IL-CIRL control scheme. A complex batch reaction system with nonlinear dynamic characteristics is considered, which includes two consecutive reactions: $A \xrightarrow{k_1} B \xrightarrow{k_2} C$. Reaction product B is the target product, while reaction product C is a by-product of batch production. Reactant A is initially added to the reactor. As the reaction proceeds, the system releases a large amount of heat. To maintain the forward progress of the reaction and increase economic benefits, cooling water is introduced to take away the heat generated during the reaction through a jacket. Specifically, a set of ordinary differential equations (ODEs) is used to model this nonlinear dynamic exothermic reaction process, and simulations are carried out according to the assumed conditions. To simplify calculations without losing generality, it is assumed that the reactants are completely mixed in the reaction vessel that has good thermal insulation, and the heat loss is negligible. The physical parameters in the reaction system are known, as shown in Table 1.

Under the above assumptions, the mass balance equations for reactant A and target product B in this batch process are described by the following ODEs:

$$\frac{dC_A}{dt} = -k_1 C_A^2 \tag{46}$$

$$\frac{dC_B}{dt} = k_1 C_A^2 - k_2 C_B \tag{47}$$

where $k_1$ and $k_2$ are temperature-related reaction rate constants, calculated as follows:

$$k_1 = \alpha_1 e^{-E_1/RT} \tag{48}$$

$$k_2 = \alpha_2 e^{-E_2/RT} \tag{49}$$

The energy balance equations for the reactor and the jacket can be derived as follows:

$$\frac{dT}{dt} = \frac{-\lambda_1}{\rho C_P} k_1 C_A^2 - \frac{-\lambda_2}{\rho C_P} k_2 C_B - \frac{Q_J}{V \rho C_P} \tag{50}$$

$$\frac{dT_J}{dt} = \frac{F_{W0}}{V_J}(T_{J0} - T_J) + \frac{Q_J}{C_J V_J \rho_J} \tag{51}$$

The heat exchange between the jacket and the reactor can be written as follows:

$$Q_J = h_{ow} A_o (T - T_J) \tag{52}$$

In addition, at the initial time step of each batch, the four system states are set as follows:

$$C_A = 1 \text{mol/L}, \ C_B = 0 \text{mol/L}, \ T = T_J = 323 \text{K} \tag{53}$$

Reasonable upper and lower limits should be provided for the control action. According to the physical constraints of the cooling water flow rate, its operating range is set as:

$$0 \text{L/s} \leq F_{ow} \leq 10 \text{L/s} \tag{54}$$

The reaction temperature, as a key parameter for batch process operation, is limited to the following range based on practical conditions:

$$298 \text{K} \leq T \leq 378 \text{K} \tag{55}$$

In the experiment, the terminal time of each batch operation is fixed as $T_f = 1\text{h}$. To better simulate the scenario where real-time product quality is undetectable in actual industrial sites, the concentrations of reactant A ($C_A$) and target product B ($C_B$) can only be obtained at the terminal time; it is assumed that other observation variables can be obtained online in real time.

### 4.1.2 Optimization control objectives

The entire control system consists of two parts: the RTO steady-state optimization layer and the IL-CIRL control layer. The goal of the RTO layer is to maximize economic benefits and the quality of the final product through the economic optimization, and the solved optimal nominal trajectory is transmitted to the IL-CIRL control layer for execution. The economic objective function used in this experiment is calculated as follows:

$$\min J_{RTO} = \min \left[ (\mathbf{C}_B(T_f) - C_{B,sp})^2 V + k \sum_{i=0}^{T_f - \Delta T} \mathbf{F}_{ow}^2(t) \right] \tag{56}$$

where $\mathbf{C}_B(T_f)$ is the product concentration at the end of the batch, $C_{B,sp}$ represents the desired target product concentration, $\mathbf{F}_{ow} = [\mathbf{F}_{ow}(0) \cdots \mathbf{F}_{ow}(T_f - \Delta T)]$ is the cooling water flow rate of the entire batch and $\Delta T$ is the sampling interval. The first term of the economic objective function in Eq.(56) is the terminal cost of batch process optimization control, indicating that the closer the concentration of target product B is to the target value, the higher the economic benefit; the second term represents the operating cost, meaning that the lower the flow rate of cooling water, the better the economic benefit. $C_{B,sp} = 0.58 \text{mol/L}$ is the target value, and $k = 0.05$ is the unit cost of the cooling water. Based on the above constraints, by optimizing this objective function, the optimal nominal reference trajectory can be obtained, including temperature $\mathbf{T}_{nom}$, concentrations $\mathbf{C}_{A,nom}$ and $\mathbf{C}_{B,nom}$, and flow rate $\mathbf{F}_{ow,nom}$.

For ILC-based batch process economic optimization control, the objective function can directly inherit the economic objective from the RTO layer, which is calculated as follows:

$$\min J_{EMPC} = \min \left[ (\Delta \mathbf{C}_B(T_f) + \mathbf{C}_{B,nom}(T_f) - C_{B,sp})^2 V + k \sum_{i=0}^{T_f - 1} (\Delta \mathbf{F}_{ow}(t) + \mathbf{F}_{ow}(t))^2 \right] \tag{57}$$

where $\Delta \mathbf{C}_B(T)$ and $\Delta \mathbf{F}_{ow}$ represent the variation of the concentration of product B and the cooling water flow rate, respectively.

### 4.1.3 IL-CIRL agent & algorithm setting

The IL-CIRL algorithm designs the agent framework based on the PPO algorithm. Specifically, the agent consists of two parts: a critic network and an actor network, with corresponding parameters set. The critic network takes the actual industrial process variables $\hat{x}$ filtered by the Kalman filter as input. Its network structure is composed of two fully connected layers, each containing 100 neurons and using the Tanh activation function. Finally, a fully connected layer outputs results corresponding to the number of actions. The input of the actor network is also the aforementioned state input, which is processed through two fully connected layers (each with 100 neurons and using the Tanh activation function); the mean path first passes through a fully connected layer, then is processed by the Tanh activation function and a scaling layer, so that the output range is (0, 10).

During pre-training, the action $a_t$ (i.e., cooling water flow rate) generated by the actor network does not directly interact with the actual industrial process but conducts preliminary policy learning updates through a simulated environment. During online training, the action $a_t$ generated by the actor network is weight-fused with the action from the ILC controller before interacting with the actual industrial control system. This interaction yields a new state $s_{t+1}$ and a calculated reward $r_t$, which are used to iteratively update the control policy.

The hyperparameters of IL-CIRL and PPO algorithms used in both pre-training and online training are shown in Table 2.

Table 2. Hyperparameters of IL-CIRL and PPO algorithms

| Hyperparameters | Values |
| --- | --- |
| critic learning rate | 1e-4 |
| actor learning rate | 5e-5 |
| number of epochs | 10 |
| discount factor | 0.99 |
| entropy loss weight | 0.02 |
| minibatch size | 64 |
| experience horizon | 2048 |
| clip factor | 0.2 |

### 4.1.4 State, action, and reward

According to the design concept of IL-CIRL described in Section 3 and combined with the characteristics of the batch reaction object, the state in MDP is defined as $T$, $T_J$, $C_A$, and $C_B$, which are the reaction temperature, jacket temperature, concentration $C_A$, and concentration $C_B$ respectively. On the other hand, the action $a_t$ in MDP is defined as the control input, specifically the cooling water flow rate, and its value range is between [0,10].

In the IL-CIRL framework, the reward function in the pre-training stage takes the output difference between the ILC controller and the RL controller as the criterion, and its expression is as follows:

$$r_t = -|u_{ILC} - u_{RL}| \qquad (58)$$

The reward function for online training is divided into continuous reward and discrete reward. Considering the advantages of hybrid reward signals in the design of setpoint tracking controllers, usually, continuous reward signals can improve the convergence during the training process, while discrete reward signals help guide the agent towards more favorable regions in the state space. Specifically, the total reward $r_t$ is the sum of the continuous reward and the discrete reward, i.e.,

$$r_t = r_c + r_d \tag{59}$$

where $r_c$ is the continuous reward, defined as:

$$r_c = \alpha \cdot |u_{ILC} - a_t| \tag{60}$$

And $r_d$ is the discrete reward, which is related to the absolute value of the error signal. Therefore, the discrete reward signal aims to drive the RL agent to move stably around the reference trajectory. The settings for discrete rewards in the experiment are given in Table 3.

**Table 3 Discrete reward settings in the experiment**

| $|x - x_{ref}|$ | < 0.05 | < 0.1 | < 0.5 | < 1 | < 2 | < 3.5 | < 5 | ≥ 5 |
|---|---|---|---|---|---|---|---|---|
| $r_d$ | 300 | 100 | 50 | 0 | -5 | -20 | -50 | -100 |

## 4.2 Control performance and convergence of the Kalman filter-based hierarchical ILC informer

First, the steady-state RTO is used to optimize the economic objective function in Eq.(56), and the optimal nominal reference trajectory of the batch process is obtained. Since the steady-state optimization of RTO does not consider fast-changing information such as disturbances, the obtained trajectory is only the optimal reference under ideal conditions. To verify the effectiveness of the batch-to-batch and within-batch hierarchical ILC based on the Kalman filter in dealing with multi-source disturbances such as deterministic and uncertain ones, the cooling water temperature in the control layer is changed from the nominal value $T_{J0} = 323K$ to the actual value $T_{J0} = 318K$, i.e., there is an offset of $-5K$ in the cooling water inlet temperature. The two types of random disturbances in Eqs.(8)-(9) are set as $\mathbf{v}_k \sim N(\mathbf{0}, 0.3)$ and $\mathbf{w}_k \sim N(\mathbf{0}, 0.4)$, respectively; the measurement noises of the observation variable and quality variable in Eq.(2) are set as $\mathbf{m}_k \sim N(\mathbf{0}, 0.06)$ and $\mathbf{n}_k \sim N(\mathbf{0}, 0.005)$, respectively.

By using the aforementioned batch-to-batch and within-batch ILC algorithm based on Kalman filtering for state estimation and iterative learning, Figs. 4-7 present the dynamic evolution results of the state trajectories of ILC during the iterative learning process as the batch number changes, showing the three-dimensional views of the four states $\mathbf{C}_A$, $\mathbf{C}_B$, $\mathbf{T}$ and $\mathbf{T}_J$ of the batch process respectively. Fig. 8 presents the evolution result of the action trajectory $\mathbf{F}_{ow}$ of ILC during the iterative learning process as the batch number changes. According to the experimental results, it can be seen that as the batch process ILC undergoes continuous iterative updates, the system states and actions gradually converge to a relatively small range, which indicates that the batch-to-batch and within-batch ILC algorithm based on Kalman filtering can achieve asymptotically convergent control of the batch process.

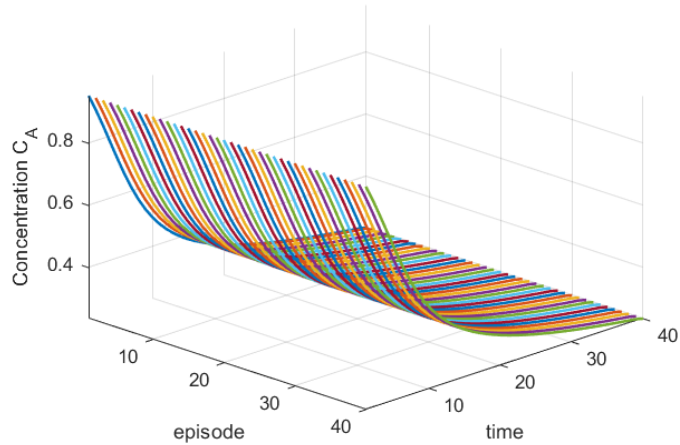

Fig. 4 3D Visualization of the iterative learning process for the concentration-A trajectory using Kalman filter-based ILC

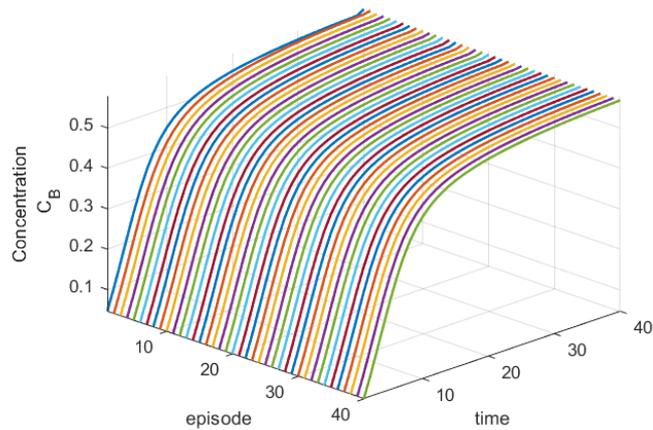

Fig. 5 3D Visualization of the iterative learning process for the concentration-B trajectory using Kalman filter-based ILC

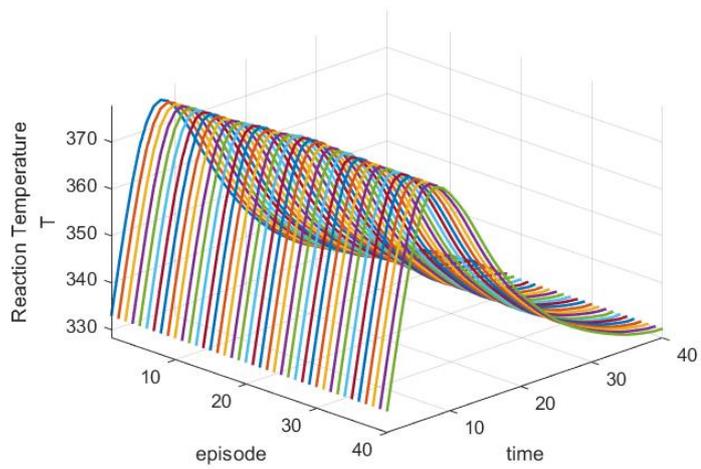

Fig. 6 3D Visualization of the iterative learning process for the reaction temperature trajectory using Kalman filter-based ILC

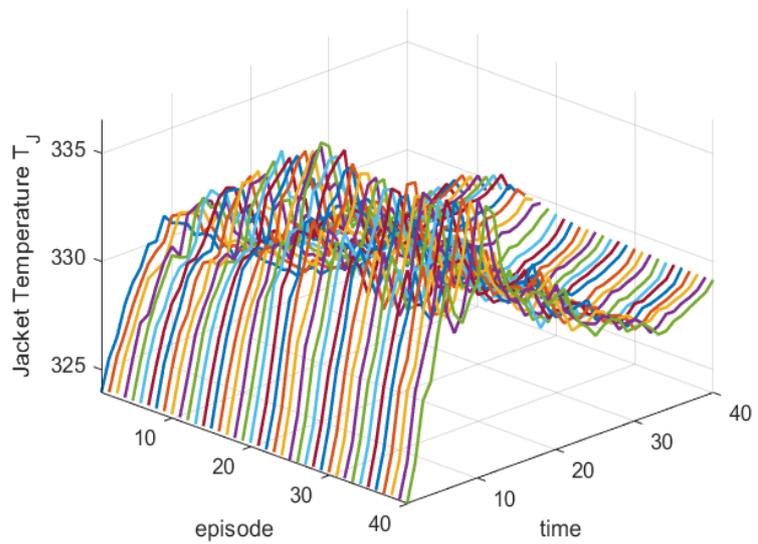

Fig. 7 3D Visualization of the iterative learning process for the jacket temperature trajectory using Kalman filter-based ILC

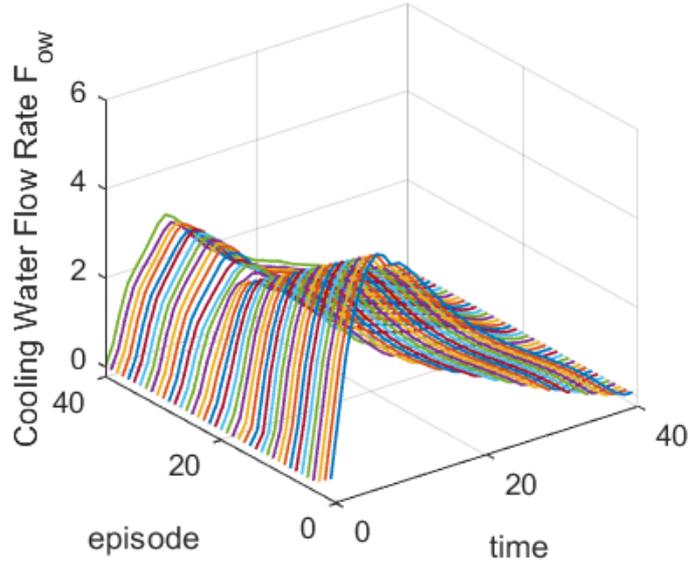

Fig. 8 3D Visualization of the iterative learning process for the action trajectory using Kalman filter-based ILC

### 4.3 The pre-training process and convergence of the proposed IL-CIRL algorithm

Table 4 Hyperparameters of the IL-CIRL and PPO algorithms

| Hyperparameters | Values |
| --- | --- |
| critic learning rate | 1e-4 |
| actor learning rate | 5e-5 |
| number of epochs | 10 |
| discount factor | 0.99 |
| entropy loss weight | 0.02 |
| minibatch size | 64 |
| experience horizon | 2048 |
| clip factor | 0.2 |

After verifying the convergence and safety of the Kalman filter and ILC iterative learning process, the designed ILC control law is used to guide the offline pre-training of the IL-CIRL controller. It is important to emphasize that the offline pre-training process does not require interaction with the actual batch process; instead, it interacts with the Kalman filter state estimates and the corresponding ILC control law. Specifically, the Kalman filter updates the system state estimate at each time step of the RL training episode, and the latest estimated state is assigned to the RL agent as the state for the next time step. This enables purely offline pre-training without any safety hazards. Additionally, the core of the IL-CIRL agent is implemented using the classical PPO algorithm, with specific training hyperparameters listed in Table 4.

Figs. 9-10 present 3D visualizations of the iterative process of state and action trajectories of IL-CIRL during offline pre-training. It can be observed that in the initial training episodes, the error between IL-CIRL and the reference trajectory is relatively large. As the Kalman filter's state estimation becomes increasingly accurate and the ILC iterative process achieves asymptotic convergence, the RL controller obtained by IL-CIRL also tends to converge.

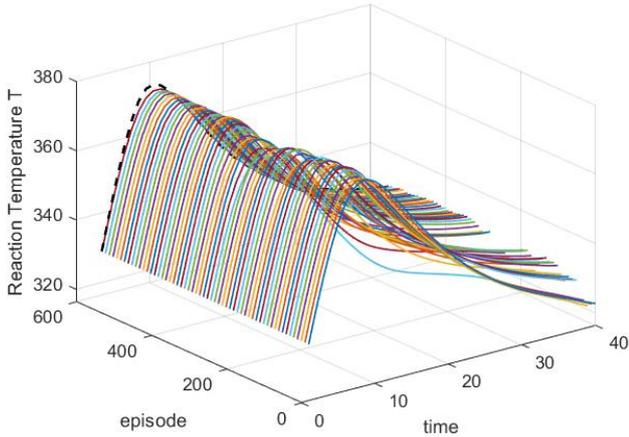

Fig. 9 3D Visualization of the iterative learning process for the reaction temperature trajectory during offline pre-training in the IL-CIRL framework

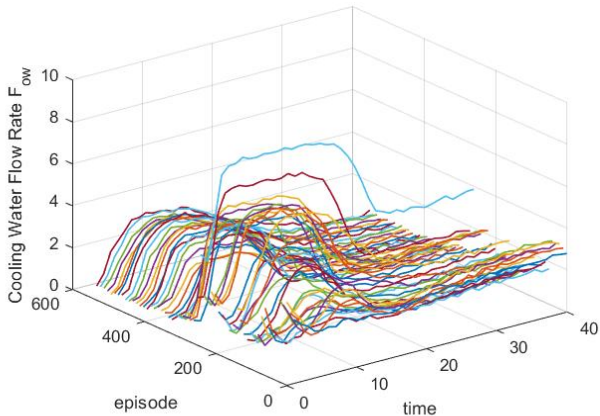

Fig. 10 3D Visualization of the iterative learning process for the action trajectory during offline pre-training in the IL-CIRL framework

**4.4 The weighted training process and convergence of the proposed IL-CIRL algorithm**

As outlined in Section 3.2, the offline pre-training process does not involve interaction with the actual variable-mode batch process. In fact, the IL-CIRL agent does not participate in the control of the batch process during this stage; its sole purpose is to enable the IL-CIRL agent to initially learn from the iterative controller, thereby improving the efficiency of the subsequent weighted training process. In the weighted training process, the control action interacting with the environment is synthesized by the iterative controller and the IL-CIRL agent according to a weight coefficient. Moreover, the state of the IL-CIRL agent consists of environmental parameters filtered by the Kalman filter. During training, based on the reward settings, the control policy of the IL-CIRL agent is further refined. Meanwhile, the weight coefficient is automatically adjusted to limit the influence of irrational behaviors of the IL-CIRL agent on the control action, ensuring the safety of practical implementation. Figs.11-13 present 3D visualizations of the weight variation, action trajectory, and state trajectory of IL-CIRL during the weighted training process. It can be observed that in the initial training episodes, the iterative controller accounts for a large proportion of the control action. As the training episodes progress, the weight gradually decreases, and the control performance gradually converges to be close to the reference trajectory.

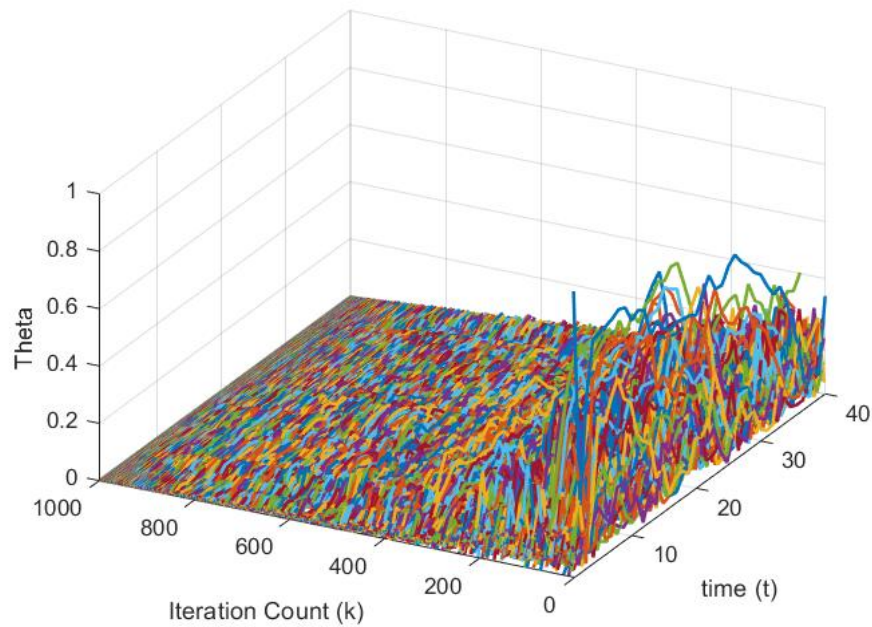

Fig. 11 Weight iteration graph of the IL-CIRL algorithm, including the weight training process

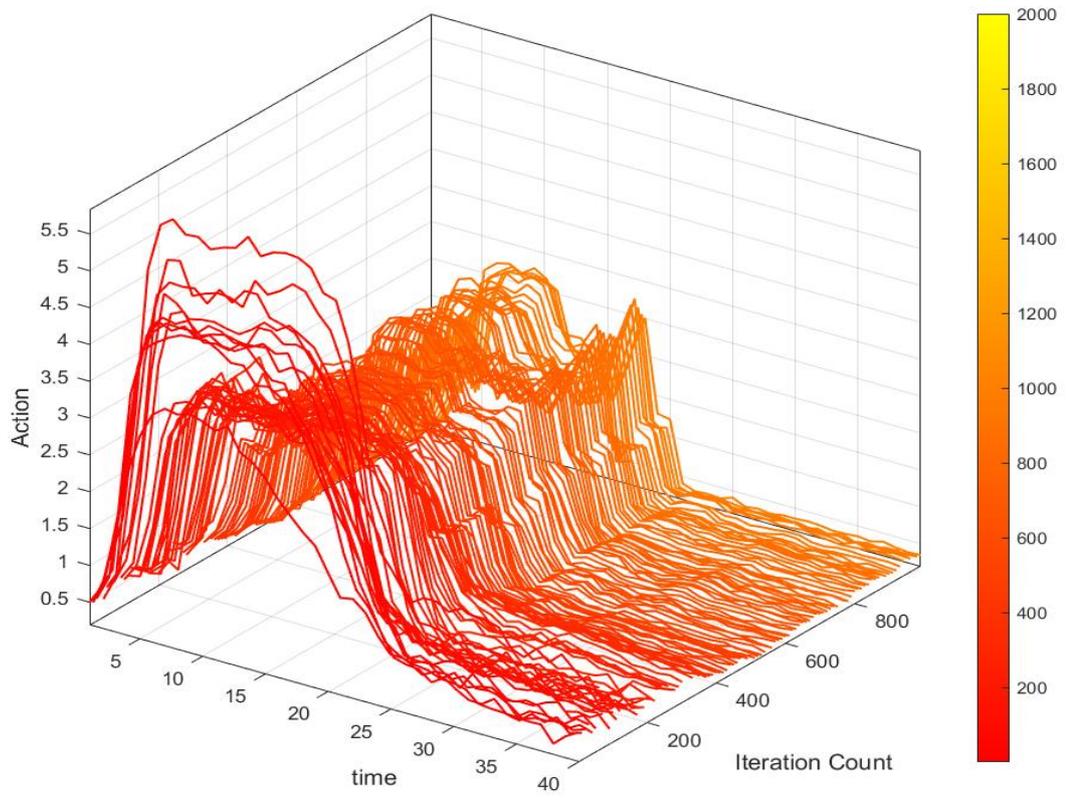

Fig. 12 3D Visualization of the iterative learning process for the action trajectory in the IL-CIRL algorithm

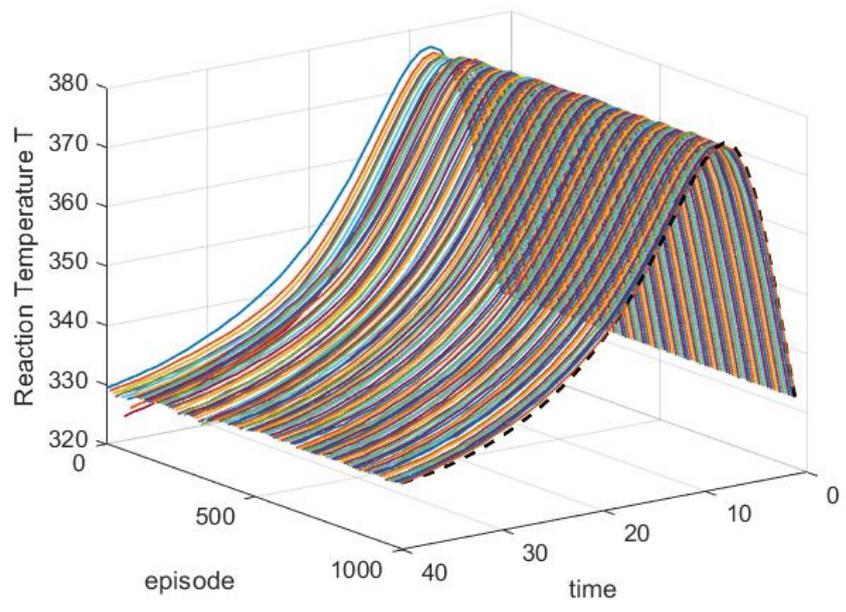

Fig. 13 3D Visualization of the iterative learning process for the reaction temperature trajectory in the IL-CIRL algorithm

## 4.5 Comparison of IL-CIRL and original PPO algorithms

Table 5 Hyperparameters of IL-CIRL and PPO algorithms

| Hyperparameters | Values |
|---|---|
| critic learning rate | 1e-4 |
| actor learning rate | 5e-5 |
| number of epochs | 10 |
| discount factor | 0.99 |
| entropy loss weight | 0.02 |
| minibatch size | 64 |
| experience horizon | 2048 |
| clip factor | 0.2 |

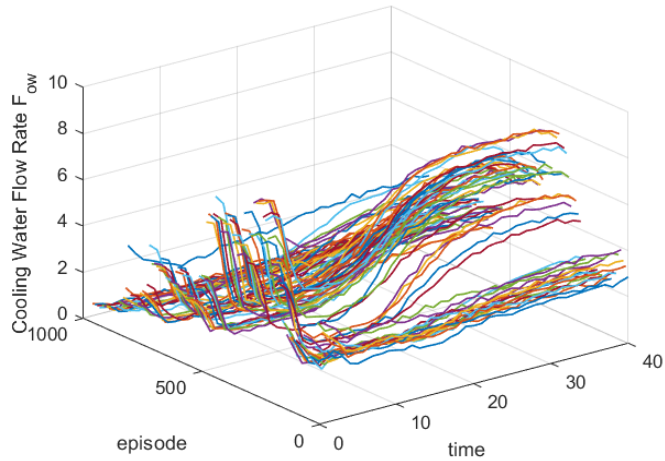

Fig. 14 Action iteration graph of the original PPO algorithm

To demonstrate that the RL agent guided by control information exhibits higher safety and training efficiency compared to directly trained RL agents, the iterative learning performance of the IL-CIRL algorithm is compared with that of the classical PPO algorithm (detailed in Section 2). The control curves and control performance changes of the classical PPO algorithm during iteration are shown in Fig. 14 and Fig. 15. Fig. 16 presents the variation curves of the Mean Squared Error (MSE) between the state trajectories of the two algorithms and the reference trajectory during iteration. It should be noted that all hyperparameters of IL-CIRL and PPO are set identically to ensure a fair comparison, with specific training hyperparameters listed in Table 5.

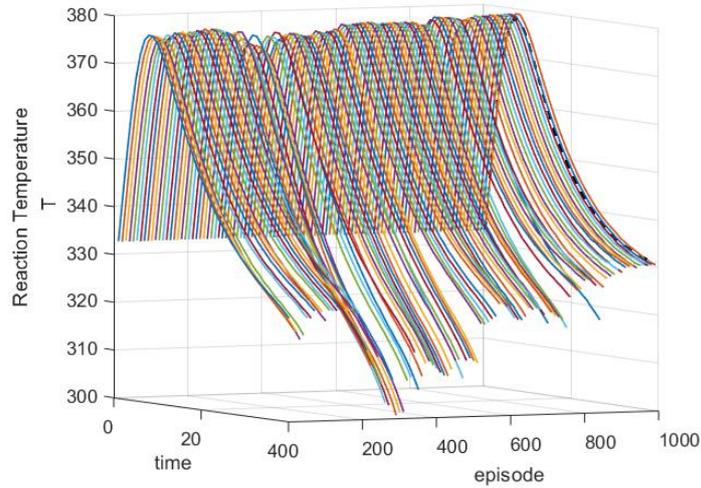

Fig. 15. Reaction temperature trajectory iteration graph of the original PPO algorithm

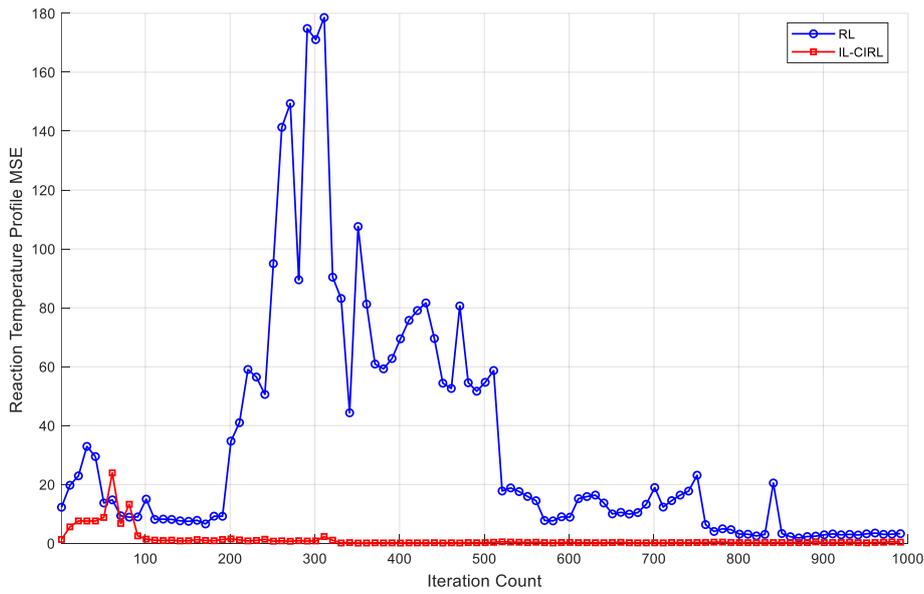

Fig. 16 Comparison of mean squared error (MSE) between RL and IL-CIRL

From the experimental results, it can be observed that the IL-CIRL algorithm achieves relatively good control performance from the initial iteration of online implementation, whereas the original PPO algorithm exhibits poor control performance in the early stages. Furthermore, as the actual batch process evolves iteratively, the control performance of PPO never surpasses that of the proposed IL-CIRL method. Therefore, by using ILC and process disturbance information to guide the pre-training process, this study effectively achieves safe offline pre-training and fast online transfer learning. This minimizes the significant safety risks posed by the trial-and-error nature of traditional DRL algorithms to the optimal control of complex industrial objects (represented by continuous batch processes).

## 5 Conclusion

This study proposes an IL-CIRL framework that integrates ILC and process disturbance information guidance. The framework aims to address safety hazards and convergence issues that RL may face during interaction with industrial plants. The key idea is to combine ILC with RL, leveraging the state estimation based on the Kalman filter and batch-to-batch/within-batch optimization control in ILC to guide the training process of the RL controller. This effectively avoids potential risks caused by random exploration. Specifically, the IL-CIRL framework provides a stable control strategy through iterative learning and ensures the safety, stability, and asymptotic convergence of the control process by real-time estimation of process disturbances and accurate state feedback. Experimental results verify the effectiveness of the method in adaptive optimal control under variable-mode environments with multi-source disturbances. This approach offers a novel solution for the optimal control of batch processes, highlights the role of control-informed guidance information in the RL training process, and exhibits broad application prospects in industrial automation.